\documentclass[aps,pra,onecolumn]{revtex4-1}

\usepackage{verbatim}
\usepackage{dsfont} 
\usepackage{amsmath}
\usepackage{amsthm}
\usepackage{graphicx}
\usepackage{mathtools} 
\usepackage{amssymb} 
\usepackage{bm}
\usepackage{enumerate}
\usepackage{tcolorbox}
\usepackage{hyperref} 
\hypersetup{colorlinks=true, linkcolor=blue, citecolor=blue, unicode=true}

\newcommand{\ket}[1]{\left| #1 \right\rangle}

\newcommand{\ketbra}[2]{\left|#1 \rangle \langle #2 \right|}

\begin{document}

\title{
	Emergence of maximal hidden quantum correlations and its trade-off with the filtering probability in dissipative two-qubit systems
}

\author{Andr\'es F. Ducuara$^{1,2,3,4,5}$} 
\email[]{andres.ducuara@bristol.ac.uk}

\author{Cristian  E. Susa$^{1,6}$}
\email[]{cristiansusa@correo.unicordoba.edu.co}

\author{John H. Reina$^{1,7}$}

\affiliation{$^{1}$Centre for Bioinformatics and Photonics---CIBioFi, Calle 13 No.~100-00, Edificio 320 No.~1069, 760032 Cali, Colombia \looseness=-1}

\affiliation{$^{2}$ Quantum Engineering Centre for Doctoral Training, University of Bristol, Bristol BS8 1FD, United Kingdom \looseness=-1} 

\affiliation{$^{3}$Quantum Engineering Technology Labs, University of Bristol, Bristol BS8 1FD, United Kingdom \looseness=-1}

\affiliation{$^{4}$H.H. Wills Physics Laboratory, University of Bristol, Tyndall Avenue, Bristol, BS8 1TL, United Kingdom \looseness=-1}

\affiliation{$^{5}$Department of Electrical and Electronic Engineering, University of Bristol, Bristol BS8 1FD, United Kingdom \looseness=-1}

\affiliation{$^{6}$Departamento de F\'isica y Electr\'onica, Universidad de C\'ordoba, 230002 Monter\'ia, Colombia\looseness=-1}

\affiliation{$^{7}$Departamento de F\'isica, Universidad del Valle, 760032 Cali, Colombia \looseness=-1}

\date{\today}

\begin{abstract}
We investigate the behaviour of quantum CHSH-nonlocality, $\rm F_3$-steering, and usefulness for teleportation in an interacting two-qubit dissipative system. We show regimes where these three quantum correlations can be extracted by means of local filtering operations, despite them not being displayed in the bare natural time evolution. Moreover, we show the existence of local hidden state (LHS) and local hidden variable (LHV) models for some states during the dynamics and thus, showing that apparently-useless physical systems could still exhibit quantum correlations, which are hidden from us, but that can still be revealed by means of local filtering operations and therefore, displaying the phenomenon of \emph{hidden} quantum correlations. Furthermore, we report on extreme versions of these phenomena, where the revealed correlations achieve the maximal amount allowed by quantum theory. This phenomenon of \emph{maximal} hidden correlations relies on the qubits collective damping, and may take place even in long-distance separated qubits. Despite the immediate appeal of the physical system displaying such an extreme phenomenon, we furthermore show however, that there actually exists a trade-off between the amount of quantum correlations which can be extracted and the filtering probability with which such protocol can be implemented. Explicitly, the higher the amount of correlations to be extracted, the more difficult it becomes for the protocol to be implemented (lower filtering probability). This is consequently showing us the remarkable fact that whilst the phenomenon of maximal hidden quantum correlations does naturally emerge during the evolution of physical systems, Nature does not completely give it away for free, by imposing a limit to the rate at which this can be done. From a theoretical point of view, the existence of such trade-off imposes a fundamental limit to the extraction of quantum correlations by local filtering operations. From a practical point of view on the other hand, the results here presented determine the amount of resources that should be invested in order to extract such maximal hidden quantum correlations.
\end{abstract}

\maketitle
\section{Introduction}

The study of non-classical correlations has remained an intriguing aspect of the theory of quantum mechanics almost since the formalisation of the theory itself. Even though it initially started with the exploration of properties like entanglement~\cite{review_entanglement} and Bell-nonlocality~\cite{review_nonlocality, book_nonlocality}, it is nowadays clear that there is actually a plethora of different non-classical phenomena including: EPR-steering~\cite{review_steering1, review_steering2}, contextuality~\cite{review_noncontextuality}, discord~\cite{review_discord}, coherence~\cite{review_coherence}, amongst many others~\cite{review_correlations}. Recently, the study of these properties has found a renewed interest under the unifying umbrella of quantum resource theories~\cite{review_QRTs}. 

One of the appealing characteristics of these counter-intuitive quantum properties is that they can be exploited for implementing operational tasks like device-independent and semi device-independent information-processing protocols~\cite{review_device_independent1}. This realisation was a decisive point that helped establish the so-called second quantum revolution~\cite{review_2QR,  review_device_independent2}. As a consequence of this, the identification and classification of quantum properties of physical systems, in addition to being of a purely theoretical interest, it also became of crucial importance for the development of quantum technologies~\cite{Pan20,quantumip2019}. These quantum properties are however fragile when considering real physical systems due to their inevitable interaction with a surrounding environment, a fact that translates into a loss of vital information (e.g., quantum correlations) stored in the relative phases (coherence) of their quantum interference dynamics, in a process known as decoherence~\cite{book_NC,weiss,breuer}. Although this usually occurs as an exponential decay, the peculiar behaviour of sudden death (decaying to zero at a finite time) was discovered for entanglement \cite{yu2009}, with other interesting behaviours such as entanglement death-revival, and sudden birth also being reported \cite{review_open_systems,SR2010, SD_SB_discord, EXP_OQS1, EXP_OQS2, karpat}.

In order to counteract the effects of decoherence, one could think of implementing strategies to  \emph{preserve}, or more interestingly, to \emph{recover} quantum correlations. It is well-known that this is possible in particular by means of \emph{Stochastic Local Operations with Classical Communication} (SLOCC) or, more colloquially,  \emph{local filtering} operations. Local filtering was first used by Popescu~\cite{LF1} as a technique for  proving the existence of the phenomenon of \emph{hidden Bell-nonlocality}. Since then, the technique has been further explored for additional properties~\cite{LF1, LF2, LF3, LF4, LF5, LF6, LF7, LF8, LF9, f1, f2, Susa2017, n1, n2}, and it has successfully been implemented experimentally as well~\cite{EXP0, EXP1, EXP2, LF8, EXP4, EXP5}. In the particular case of entanglement, it has been proven that there are physical scenarios in which local filtering can mitigate the effect of the environment~\cite{isasi1}, and sometimes even help to recover the initial amount of entanglement~\cite{isasi2}. In this regard, the behaviour of additional properties besides entanglement, as well as their control and optimisation in physical systems remain of great research interest \cite{n1, n2, LF8, LF9}.

In this work, we show that dissipative quantum systems can reveal hidden quantum correlations for which the \emph{maximal} values allowed by quantum theory can be reached; in particular, we explicitly show that an open bipartite qubit system under collective decoherence can exhibit the phenomenon of \emph{maximal hidden quantum correlations}, one example state of such a nature is the qubit-qutrit known as the \emph{erasure state}~\cite{erasure, genuine}. We address here as \emph{quantum correlations} the property of entanglement, and three \emph{entanglement-based} properties\footnote{Quantum entanglement is considered as a counter-intuitive correlation (without classical counterpart) between quantum systems~\cite{review_correlations}. 
 On the other hand, Bell-nonlocality is considered as a property of objects called behaviours or boxes (see e.g., \cite{popescu1994b}). Thus, nonlocal quantum correlations are revealed by the existence of Bell-nonlocality. In this work, for the sake of writing simplicity, we refer to the all studied quantum properties as to quantum correlations.} 
which respect the following strict hierarchy: Bell-nonlocality $\rightarrow$ EPR-steering $\rightarrow$ usefulness for teleportation $\rightarrow$ entanglement \cite{criterion_UfT}. Meaning that if a quantum state is Bell-nonlocal, then it is automatically EPR-steerable, but not the other way around, and similarly for the other implications and properties. Specifically, we provide death-revival and sudden birth of entanglement scenarios for which the aforementioned entanglement-based properties are not present in the bare system Markovian dynamics, but that can still be revealed by means of local filtering operations. Furthermore, that these hidden quantum correlations can in fact be maximal, meaning that they achieve the maximum amount allowed by quantum theory. In this way, in the sudden birth case in particular, the physical system in consideration displays a type of \emph{all from nothing} behaviour, both in terms of the correlations it can generate via local filtering and from the temporal evolution it undergoes.

Despite the immediate appeal of having a physical system naturally evolving to display the phenomenon of maximal hidden quantum correlations, we however furthermore show that there also exists a \emph{trade-off} between the amount of correlations that can be extracted and the efficiency of the protocol which achieves this. Thus, effectively establishing a practical fundamental limit for the amount of correlations that can realistically be extracted. Besides the theoretical interest that such trade-off represents, the results here determine, in a practical manner, the amount of resources that should be invested in order to extract such maximal hidden quantum correlations.

This work is organised as follows. In Section \ref{sec:standardcorr} we introduce the quantum properties of Bell-nonlocality, EPR-steering, usefulness for teleportation and entanglement, as well as their corresponding measures. In section \ref{sec:hiddencorr}, we address the enhancement of these quantum properties by means of local filtering operations, provide details on their operational implementation, and define figures of merit which we address as \emph{hidden measures}. In Section \ref{sec:opensystem} we describe the physical system used to present our results. In Section \ref{sec:result}  we establish the main results of this work. Specifically, we show that the considered dissipative system can be tailored to display maximal hidden quantum correlations, even though it naturally evolves through entangled-unsteerable (local) states. We then proceed show the existence of the trade-off between the amount of correlations which can be extracted and the efficiency of such extraction. We end up in Section \ref{sec:conclusion} with conclusions and perspectives.

\section{Quantum correlations of interest}
\label{sec:standardcorr}

We first explore the quantum correlations of interest by employing the following measures: for entanglement we consider the concurrence \cite{criterion_concurrence}, depicted as the function ${\rm C}(\rho)$. For Bell-nonlocality we consider the violation of the Clauser-Horn-Shimony-Holt (CHSH) inequality by means of a renormalisation of the Horodecki criterion \cite{criterion_CHSH}, as the function ${\rm B}(\rho)$. For EPR-steering we consider the violation of the so-called $\rm F_3$-inequality by means of a renormalisation of the Costa-Angelo criterion \cite{criterion_F3}, as the function $\rm BF_3(\rho)$. Finally, for usefulness for teleportation we use a renormalisation of the fidelity of teleportation \cite{criterion_UfT}, as the function ${\rm D}(\rho)$. We now explicitly address these functions.

\subsection{Standard Measures}

An arbitrary two-qubit state $\rho \in D(\mathds{C}^2 \otimes \mathds{C}^2)$ can be written as $\rho =\frac{1}{4}\sum _{i,j=0}^3 R_{ij}\sigma_i \otimes \sigma_j$, with the real matrix $R_{ij}={\rm Tr} \left[ (\sigma_i \otimes \sigma_j)\rho \right]$, $\sigma_0=\mathds{1}$ and $\{\sigma_i\}, i=1,2,3$, the Pauli matrices. $R$ can be written as \cite{LF6}:
\begin{align}
	R= 
	\begin{pmatrix} 
	1& \bm{b}^{\,T}\\
	\bm{a}& T
	\end{pmatrix},
	\label{eq:R}
\end{align}
with the Bloch vectors $\bm{a}=[a_i]$, $a_i= {\rm Tr} [(\sigma_i \otimes  \mathds{1})\rho]$; $\bm{b}=[b_i]$ $b_i= {\rm Tr} [(\mathds{1}\otimes \sigma_i)\rho]$, and correlation matrix $T=[T_{ij}]$, $T_{ij}={\rm{Tr}}[\rho (\sigma_i  \otimes \sigma_j)]$. We now consider some quantities of interest: the singular values of $T$ denoted, in decreasing order, as $\{s_1, s_2, s_3\}$, and the \emph{chirality} of the state as $\chi={\rm det}(T)$ \cite{Milne}. Similarly, we consider as $\{\lambda_i\}$ the square roots of the eigenvalues of the product matrix $\rho \hat \rho$, in decreasing order, with $\hat \rho= \left(\sigma_y \otimes \sigma_y \right) \rho^* \left( \sigma_y \otimes \sigma_y \right)$, $\rho^*$ the complex conjugate of $\rho$ and, $\sigma_y$ the Pauli matrix. The explicit quantifiers for the quantum correlations considered in this work read:
\begin{align}
	{\rm B}(\rho)&= 
	\max 
	\left \{0, s_1^2+s_2^2-1\right \},
	\label{eq:B}\\
	{\rm BF_3}(\rho) &=
	\max 
	\left \{0, \frac{s_1^2+s_2^2+s_3^2-1}{2} \right \},
	\label{eq:BF3}\\
	{\rm D}(\rho) &= 
	\max
	\left \{0, \frac{|s_1|+|s_2|-\chi |s_3|-1}{2}\right \},
	\label{eq:D}\\
	{\rm C}(\rho) &= 
	\max
	\{0,\lambda_1-\lambda_2-\lambda_3-\lambda_4 \}.
	\label{eq:C}
\end{align}
The maximal violation of the CHSH inequality by an arbitrary two-qubit state can be fully characterised by the Horodecki criterion \cite{criterion_CHSH}, which establishes that $\rho$ violates the CHSH-inequality if and only if $M(\rho)=s_1^2+s_2^2>1$. Similarly, the Costa-Angelo criterion \cite{criterion_F3} establishes the maximal violation of the so-called $\rm F_3$-inequality for EPR-steering by arbitrary two-qubit states and reads ${\rm F_3}(\rho)= s_1^2+s_2^2+s_3^2>1$. Considering the standard teleportation protocol \cite{UfT1993} and its characterisation for arbitrary two-qubit states \cite{criterion_UfT}, the usual functions for exploring usefulness for teleportation include: the fidelity of teleportation $F(\rho)=\frac{1}{2}\left[\frac{1}{3}N(\rho)+1 \right]$, the singlet fraction $f(\rho)=\frac{1}{4}[1+N(\rho)]$, with the function $N(\rho) = |s_1|+|s_2|-\chi |s_3|$. A state is said to be useful for teleportation when $N(\rho)>1$ or $f(\rho)>\frac{1}{2}$ or $F(\rho)>\frac{2}{3}$. The quantifiers in Eqs. \eqref{eq:B}-\eqref{eq:D} are renormalised version of the above criteria such that they can be compared at the same level of concurrence.

\section{Hidden quantum correlations}
\label{sec:hiddencorr}

In this section we address the implementation of local filtering operations on the physical system of interest, and define the figures of merit for hidden quantum correlations.

\subsection{Description of the general protocol}

We now describe the effect of local filtering operations when potentially being implemented at each point of the dynamics. Observers Alice and Bob are interested in maximising the quantum correlations of the state $\rho(t)$ ($\rho$ during this section) by means of local filtering operations which effectively transform the state $\rho$ into a state $\rho'$. In  so doing, Alice and Bob share the two-qubit state $\rho$ and consider local binary measurements $E_W=\{E^0_W, E^1_W\}$, $W\in \{A,B\}$ and $E^0_W=f_W^{\dagger}f_W$, $E^1_W=\mathds{1}-E^0_W=g_W^\dagger g_W$, with the  complex matrices $\{f_W\}$ satisfying $f_W^{\dagger} f_W \leq \mathds{1}$. We can check that the $E_W$'s are valid local binary POVMs; $E^i_W\geq 0, \sum_i E^i_W=\mathds{1}, W\in \{A,B\}$.  The matrices $\{f_W\}$ and $\{g_W\}$ can be thought as the associated Kraus operators so that after measuring the system, the four possible unnormalised post-measured states can be determined by the measurement outcome and are given by:
\begin{align*}
	&00 \hspace{0.7cm} \longrightarrow \hspace{0.7cm}
	\tilde \rho_{00}=
	\tilde \rho_{ff}=
	(f_A \otimes f_B)\rho (f_A\otimes f_B)^{\dagger}
	,
	\\
	&01 \hspace{0.7cm} \longrightarrow \hspace{0.7cm}
	\tilde \rho_{01}=
	\tilde \rho_{fg}=
	(f_A \otimes g_B)\rho (f_A\otimes g_B)^{\dagger}
	,
	\\
	&10 \hspace{0.7cm} \longrightarrow \hspace{0.7cm}
	\tilde \rho_{10}=
	\tilde \rho_{gf}=
	(g_A \otimes f_B)\rho (g_A\otimes f_B)^{\dagger}
	,
	\\
	&11 \hspace{0.7cm} \longrightarrow \hspace{0.7cm} 
	\tilde \rho_{11}=
	\tilde \rho_{gg}=
	(g_A \otimes g_B)\rho (g_A\otimes g_B)^{\dagger}.
\end{align*}
Depending on the measurement outcome they obtain, Alice and Bob can therefore determine the post-measured state, allowing them to effectively implement a filtering process. For instance, when the measurement outcome is $00$, the realisation of the state $\tilde \rho_{ff}$ is guaranteed, which occurs with probability $p_{ff}={\rm Tr}[(f_A^\dagger f_A \otimes f_B^\dagger f_B)\rho]$. Alice and Bob would repeat the experiment enough times so to keep the post-measured state only when it has ended in the target state ($\tilde \rho_{ff}$), thus effectively implementing a local filtering process. 
A general local filtering operation can then be written as $\rho'=\frac{1}{p_{ff}}(f_A \otimes f_B) \rho (f_A \otimes f_B)^\dagger$, with $f_A$ and $f_B$ arbitrary positive semidefinite operators satisfying $f_W^{\dagger}f_W\leq \mathds{1}$. The operators $f_A, f_B$ are termed \emph{local filters} and the procedure as a whole as \emph{Stochastic Local Operations with Classical Communication} (SLOCC) or simply local filtering.

\subsection{Details on the operational implementation of arbitrary local filters}

\begin{figure}[h!]
    \centering
    \includegraphics{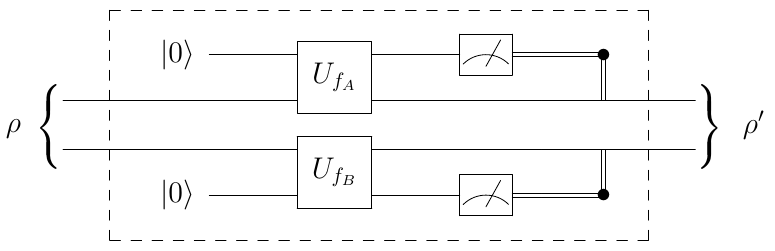}
    \caption{Circuit diagram for implementing arbitrary SLOCC transformations on two-qubit states.}
    \label{fig:fig5}
\end{figure}

We now address some details on the operational implementation of local filters, and highlight the importance of the filtering probability $p_{ff}={\rm Tr}[(f_A^\dagger f_A \otimes f_B^\dagger f_B)\rho]$ for a given state $\rho$ and local filters $f_A, f_B$. In \autoref{fig:fig5} we address a circuit implementing arbitrary local filtering operations.

The circuitry inside the dashed box in \autoref{fig:fig5} represents the implementation of an arbitrary SLOCC, which has as input a general two-qubit state $\rho$, and as output another two-qubit state $\rho'$. An SLOCC is equivalent to doing two things: {\bf i)} implementing local binary POVM measurements on Alice and Bob sides and {\bf ii)} being able to post-select a target state.

Regarding the local POVMs, Naimark's theorem \cite{wilde2013} establishes that any POVM can be implemented by considering: {\bf (a)} an ancillary system with an ancillary input state, {\bf (b)} an unitary matrix acting on both ancillary and system of interest, and {\bf (c)} a projective measurement which crucially, only acts on the ancillary state. In our particular case, the ancillary system is a one-qubit system, the ancillary input state is $\ket{0}$, the projective measurement is the one performed in the computational basis $\{\ketbra{0}{0}, \ketbra{1}{1}\}$, and the unitary matrices $U_{f_W}, W \in \{A, B\}$ which are calculated in terms of the $f_W$ are given by \cite{wilde2013}:
\begin{align}
    U_{f_W} =f_W \otimes \mathds{1}
    + \left( \sqrt{\mathds{1}-f_W^\dag f_W} \right) \otimes (-i \sigma_2), \hspace{0.3cm} \sigma_2=\begin{pmatrix}0&-i\\i&0\end{pmatrix}, \hspace{0.3cm}  W\in\{A, B\},
    \label{eq:UfA}
\end{align}
where the second subsystem is the system of the ancillary, and the matrices $f_W , W \in \{A, B\}$ are the local filters. With these elements in place, Naimark's theorem then allows for the implementation of the local filters $f_A, f_B$ as shown in \autoref{fig:fig5}.

Regarding the post-selection stage, Alice and Bob read the measurement outcomes on the ancillary system, and whenever they obtain any of the three outputs $\{ 01, 10, 11 \}$, they decide \emph{not} to use the state and repeat the experiment. They are only going to keep the state whenever they obtain the output $00$, which in turn signals that they are having the desired output target state $\rho'=\rho_{ff}$. Upon success, they can then further use the output state $\rho'$ as a resource for violating a Bell-inequality, steering-inequality, a teleportation protocol, or any other information-theoretic protocol. We can then see here that the filtering probability $p_{ff}$ is playing an important role, because it tells us the probability of arriving to the desired target state, and it can therefore be thought of as representing the efficiency of the procedure.

\subsection{Hidden measures}

Alice and Bob now want to find appropriate local filters such that they increase the values for the different measures of quantum correlations. In particular, it turns out that non-invertible operations always decrease entanglement \cite{LF6} and therefore, we further constrain the set of local filters (in addition to $f_W^{\dagger} f_W \leq \mathds{1}$) to satisfy $f_A, f_B \in {\rm GL}(2, \mathds{C})$, the group of invertible $2\times2$ complex matrices. Furthermore, amongst all possible local filtering operations, here we address the Kent-Linden-Massar (KLM) SLOCC transformation \cite{LF5}. 
The KLM-SLOCC has the crucial property of transforming any state $\rho$ into or close to its so-called \emph{Bell-diagonal unique normal form} (which we denote as $\rho^{\rm BD}_{\rm UNF}$) \cite{LF5, LF6, LF7, n2}. The KLM-SLOCC has also been proven to be the optimal local filtering for \emph{simultaneously} maximising: i) concurrence \cite{LF6}, ii) usefulness for teleportation \cite{LF7}, and iii) the violation of the CHSH-inequality for Bell-nonlocality
~\cite{LF8}. Interestingly, it is not known whether this is also the case for the $\rm F_3$-inequality for EPR-steering. In particular, as well shall see later on, this turns out to be the case for all the dynamics here considered.

Given an arbitrary $\rho$, it is then possible to derive analytic functions for exploring the quantum correlations on $\rho^{\rm BD}_{\rm UNF}$ in terms of the original state $\rho$. We address these correlations as the hidden quantum correlations of the original $\rho$, and we now define functions describing each property as follows: i) ${\rm HB}(\rho)$ (hidden CHSH-nonlocality)  \cite{criterion_HCHSH}, ii) ${\rm HBF_3}(\rho)$ (hidden $\rm F_3$-steering)~\cite{criterion_F3,DSR2022},
iii) ${\rm HD}(\rho)$ (hidden usefulness for teleportation) \cite{criterion_HD}, and iv)  ${\rm MEC}(\rho)$ (maximum extractable entanglement) \cite{LF5}\footnote{One could be tempted to call this function a ``hidden concurrence'', however, it is not difficult to check that local filtering operations do \emph{not} create entanglement and therefore, there does not exist such a thing as revealing ``hidden entanglement'' by means of local filtering. It happens however, that even though entanglement cannot be created, the \emph{amount} of it can still increase, and the maximum amount that can be extracted in such a way has been coined as \emph{maximum extractable entanglement} and therefore, we denote the case for concurrence as ${\rm MEC}(\rho)$.}. 
Furthermore, it turns out that these  last two quantities are in fact equal, ${\rm HD}(\rho)={\rm MEC}(\rho)$ \cite{criterion_HD} and therefore, we will only address ${\rm HD}(\rho)$. We remark here that a different concept has been addressed under the name of hidden entanglement \cite{r1, r2, r3, r4, r5}. 

We now address the explicit definitions of these functions. The KLM-SLOCC acts as $\rho \rightarrow \rho^{\rm BD}_{\rm UNF}$, which in terms of the R-picture this reads as $R\rightarrow R^{\rm BD}_{\rm UNF}=\text{diag}(1,\sqrt{\nu_1/\nu_0},\sqrt{\nu_2/\nu_0},-\sqrt{\nu_3/\nu_0})$, with $\{\nu_{i=0,1,2,3}\}$ the eigenvalues of the operator $\eta R\eta R^T$ in decreasing order and $\eta=\text{diag}(1,-1,-1,-1)$ \cite{LF6,criterion_HCHSH}. If we now replace the Bell-diagonal unique normal form of an arbitrary two-qubit state in the previous functions: \autoref{eq:B}, \autoref{eq:BF3}, \autoref{eq:D}, and \autoref{eq:C}, we obtain explicit measures for hidden quantum correlations as:
\begin{align}
    {\rm HB}(\rho)
    &=
    \max 
    \left \{
    0, 
    \frac{1}{\nu_0}\left (\nu_1+\nu_2-\nu_0\right)
    \right \}
    \label{eq:HB}
    \\
    {\rm HBF_3}(\rho)
    &=
    \max 
    \left \{
    0,
    \frac{1}{2\nu_0}\left (\nu_1+\nu_2+\nu_3-\nu_0\right)
    \right \},
    \label{eq:HBF3}
    \\
    {\rm HD}(\rho)
    &=
    \max
    \left \{
    0,
    \frac{1}{2\sqrt{\nu_0}}\left(\sqrt{\nu_1}+\sqrt{\nu_2}+\sqrt{\nu_3}- \sqrt{\nu_0}\right) 
    \right \}.
    \label{eq:HD}
\end{align}
Maximum extractable concurrence becomes equal to hidden usefulness for teleportation ${\rm MEC}(\rho)={\rm HD}(\rho)$ \cite{criterion_HD}. A unified framework for quantifiers of both standard and hidden quantum correlations is addressed as part of an upcoming work \cite{DSR2022}.

\section{Open system dynamics}
\label{sec:opensystem}

We now address the physical scenario which concerns two atom-like qubits in a common reservoir. Allowing the interaction of the two qubits with this bath, coherent and incoherent effects can be covered in a Markovian description of the quantum dynamics for the system Hamiltonian $H$ ($\hbar=1$)~\cite{Agarwalbook}:
\begin{align}
	\label{eq:master}
	\dot{\rho}(t)= i\left[\rho(t),H\right]
	-\sum_{i,j=1}^{2}\frac{\Gamma _{ij}}{2}\left( \left\{\rho(t),			\sigma^{(i)}_{+}\sigma^{(j)}_{-}\right\}
     -2\sigma^{(i)}_{-}{\rho(t)} \sigma^{(j)}_{+}\right),
\end{align}
with $H=-\frac{1}{2}\omega_1\sigma_z\otimes\mathds{1}-\frac{1}{2}\omega_2\mathds{1}\otimes\sigma_z+\frac{V}{2}(\sigma_x\otimes\sigma_x+\sigma_y\otimes\sigma_y)$, where $\omega_i$ is the transition frequency of qubit $i$ and
$V$ is the dipole-dipole interaction; the spontaneous emission $\Gamma_{ii}\equiv \Gamma$, and the  collective damping $\Gamma_{i\neq j}\equiv \gamma$.
$\sigma_i$ are the Pauli matrices, and $\sigma_{+}^{(i)}=\ketbra{1}{0}$ ($\sigma_{-}=\ketbra{0}{1}$) the raising (lowering) operator in the computational basis $\{\ket{0},\ket{1}\}$ for the $i$-th qubit. We now consider the initial state of the system in an X-form as:
\begin{align}
	\rho(0)=
	\rho_X=
	\left(\begin{array}{cccc}
		a&0&0&w\\
		0&b&z&0\\
		0&z^*&c&0\\
		w^*&0&0&d
	\end{array}\right),
	\label{eq:Xstate}
\end{align}
which allows an analytical solution of the master equation \eqref{eq:master}, with the time-dependent matrix elements given by:
\begin{align}
a(t) 
	&
	= \frac{e^{-t (\gamma +3 \Gamma )}}{2
		(\gamma^2 -\Gamma^2)} \left[2 (a+d) (\gamma^2 -\Gamma^2 ) e^{t (\gamma +3 \Gamma )}\right. \nonumber\\
	&
	\left.+b \left(\Gamma ^2-\gamma ^2\right) \left(e^{2 t (\gamma
		+\Gamma )}-2 e^{t (\gamma +3 \Gamma )}+e^{2 \Gamma  t}\right)+c \left(\Gamma ^2-\gamma
	^2\right) \left(e^{2 t (\gamma +\Gamma )}-2 e^{t (\gamma +3 \Gamma )}+e^{2 \Gamma  t}\right)\right.\nonumber\\
	&
	\left.+2
	d \left((\gamma -\Gamma
	)^2 e^{2 t (\gamma +\Gamma )}-\left(3 \gamma ^2+\Gamma ^2\right) e^{t (\gamma +\Gamma )}+(\gamma +\Gamma )^2 e^{2 \Gamma  t}\right)\right.\nonumber
	\\
	&\left.-2 \left(\Gamma
	^2-\gamma ^2\right) \Re(z) \left(e^{2 \gamma  t}-1\right) e^{2 \Gamma  t}\right] ,\nonumber
	\\
	\left(\begin{array}{c}
	b(t)\\
	c(t)
	\end{array}\right) 
	&=
	\frac{1}{4 (\Gamma^2 -\gamma^2 )}\left[2 d e^{-t (\gamma +2 \Gamma )} \left(-2 \left(\gamma ^2+\Gamma ^2\right) e^{\gamma 
		t}+(\gamma -\Gamma )^2 e^{t (2 \gamma +\Gamma )}+(\gamma +\Gamma )^2 e^{\Gamma  t}\right)\right.\nonumber
		\\
	&\left.\pm2
	(\gamma^2 -\Gamma^2 ) e^{\Gamma  (-t)} \left(\mp(b+c) \cosh (\gamma 
	t)-(b-c) \cos (2 t V)\right.\right.\nonumber
	\\
	&
	\left.\left.+2 \Im(z) \sin (2 t V)\pm2 \Re(z) \sinh
	(\gamma  t)\right)\right] ,\nonumber
	\\
	z(t) 
	&=
	\frac{1}{4 (\gamma^2 -\Gamma^2 )}\left[2 d e^{-t (\gamma +2 \Gamma )} \left((\gamma -\Gamma )^2 e^{t (2 \gamma +\Gamma
		)}-(\gamma +\Gamma )^2 e^{\Gamma  t}+4 \gamma  \Gamma  e^{\gamma  t}\right)\right.\nonumber\\
	&\left.+2 (\gamma^2 -\Gamma^2
	) e^{\Gamma  (-t)} \left(-(b+c) \sinh (\gamma  t)+i
	(b-c) \sin (2 t V)\right.\right.\nonumber
	\\
	&\left.\left.+2 i \Im(z) \cos (2 t V)+2 \Re(z) \cosh
	(\gamma  t)\right)\right] ,\nonumber\\
	d(t) &= d e^{-2 \Gamma  t} ,\;\;\; w(t) = w e^{-t (\Gamma -2\omega_0 i)} ,\label{eq:analyticalsol}
\end{align}
where $\omega_0\equiv\frac{\omega_1+\omega_2}{2}$. 

The quantum correlations of the considered open system \eqref{eq:master} can then explicitly be calculated using functions \eqref{eq:B}-\eqref{eq:C} for standard measures, and \eqref{eq:HB}-\eqref{eq:HD}, for hidden measures. It turns out that although there are scenarios of death-revival and sudden birth of entanglement, this entanglement is weak, as it does not allow the revival or sudden birth of additional entanglement-based correlations. Explicitly, for the case of sudden birth of entanglement, it means that during the whole temporal evolution of the system we have: ${\rm B}(\rho(t))=0$ (Bell-nonlocality), ${\rm BF_3}(\rho(t))=0$  (EPR-steering), and ${\rm D}(\rho(t))=0$ (usefulness for teleportation). We now address the effect on the system's correlations when two observers, say Alice and Bob, implement local filtering operations, with the goal of revealing hidden quantum correlations, if present at all.

\section{Main results}
\label{sec:result}

Our main result relies on the existence of entanglement-based quantum correlations, that naturally are hidden for us, but that can be revealed and maximised by making use of local filters. This is shown for a quantum open system undergoing the Markovian dynamics leading by the solution \eqref{eq:analyticalsol}. Fixing the initial state to be $\rho_W(p)=\frac{(1-p)}{4}\mathds{1}+p\ketbra{\psi^+}{\psi^+}$, with $\ket{\psi^+} = \frac{1}{\sqrt{2}}(\ket{01}+\ket{10})$ and the mixing parameter $p\in[0,1]$, Eq. \eqref{eq:analyticalsol}  becomes independent of the strength $V$ such that the collective damping $\gamma$ holds as the relevant parameters of the dynamics. Making $k=\gamma/\Gamma$, $k\in(0,1]$, solution \eqref{eq:analyticalsol} reduces to:
\begin{eqnarray}
	&&a(t) = 1-2b(t)-d(t), \; c(t)=b(t), \; d(t) = \frac{1}{4} (1-p) e^{-2 \Gamma  t},
	\label{eq:d}\nonumber
	\\
	&&b(t) = \frac{e^{-2 \Gamma  t}}{4 \left(1-k^2\right)} \Big\{ 2 e^{\Gamma  t} \big[\left(p(k^2+k-1)-k\right) \sinh
   (\Gamma  k t) 
  +\left(1-k^2 p\right) \cosh (\Gamma  k t)\big]-\left(1+k^2\right)
   (1-p)\Big\},
   \nonumber
   \\
	&&z(t) = \frac{e^{-2 \Gamma  t}}{2 \left(1-k^2\right)} \Big\{e^{\Gamma  t}\big[\left(p\left(1-k^2\right)+(1-p)k\right)\cosh(\Gamma kt)
	-\left(1-k^2 p\right) \sinh (\Gamma kt)\big] +kp-k\Big\}.	
 	\nonumber
\end{eqnarray}

The exact formulae for the standard and hidden correlations of interest are explicitly given as shown in \autoref{table:shq}. 
\begin{table}[h]
\centering
\begin{tabular}{|l|l|l|} 
\hline
{\bf Correlation} & {\bf Standard measure}&{\bf Hidden measure}\\
\hline
CHSH-NL 
& 
$\text{B}(\rho(t))=\max{\left\{0,\left(\varepsilon(t)\right)^2+4|z(t)|^2-1\right\}}$  
&
$\text{HB}(\rho(t))=\max{\left\{0,\frac{\left(\digamma_-(t)\right)^2+|z(t)|^2}{\left(\digamma_+(t)\right)^2}-1\right\}}$
\\
\hline
$\rm F_3$-steering 
&
$\text{BF}_3(\rho(t))=\max{\left\{0,\frac{\left(\varepsilon(t)\right)^2+8|z(t)|^2-1}{2}\right\}}$
&
$\text{HBF}_3(\rho(t))=\max{\left\{0,\frac{1}{2}\left(\frac{\left(\digamma_-(t)\right)^2+2|z(t)|^2}{\left(\digamma_+(t)\right)^2}-1\right)\right\}}$
\\
\hline
Teleportation 
&
$\text{D}(\rho(t))=2\max{\left\{0,|z(t)|-\frac{b(t)+c(t)}{2}\right\}}$
&
$\text{HD}(\rho(t)) =\frac{1}{\digamma_+(t)}\max{\left\{0,|z(t)|-\sqrt{a(t)d(t)}\right\}}$
\\
\hline
Entanglement 
&
$\text{C}(\rho(t))=2\max{\left\{0,|z(t)|-\sqrt{a(t)d(t)}\right\}}$  
&
$\text{MEC}(\rho(t))=\text{HD}(\rho(t))$
\\
\hline
\end{tabular}
\caption{Analytical solution for standard (unfiltered) and hidden (filtered) quantum correlations in terms of the density matrix elements. $\varepsilon(t)=a(t)+d(t)-b(t)-c(t)$, $\digamma_{\pm}(t)=\sqrt{b(t)c(t)}\pm\sqrt{a(t)d(t)}$ and $a(t), b(t), c(t), d(t), z(t)$ as defined in the main text.}
\label{table:shq}
\end{table}

\subsection{Death and revival of maximal hidden quantum correlations}

As a first result, we report on the emergence of \emph{death and revival} of maximal hidden quantum correlations.

\begin{figure}[h!]
	\centerline{
		\includegraphics[scale=0.8]{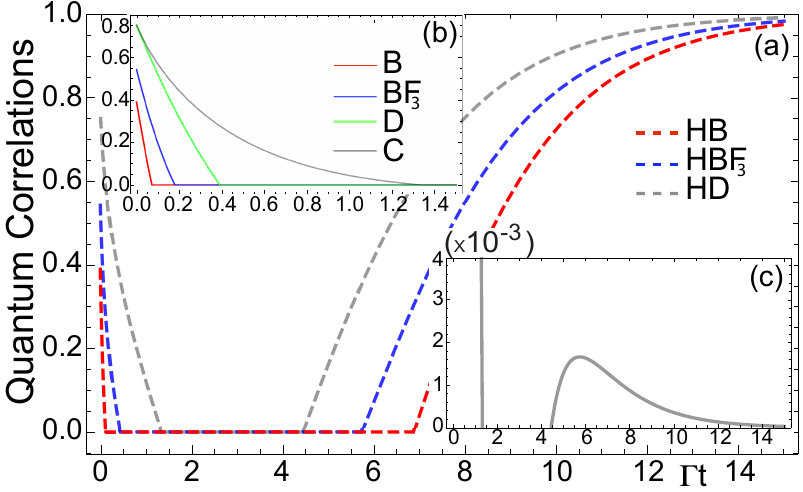} 
	}
	\caption{
	Sudden death and revival of quantum correlations. (a) Hidden quantum correlations, (b) Standard quantum correlations, (c) Concurrence death-revival. The plots show CHSH-nonlocality (red), $\rm F_3$-steering (blue), usefulness for teleportation (green) and concurrence (gray). HD$\equiv$MEC. Solid curves represent standard correlations while dashed curves stand for hidden correlations. Collective damping  $\gamma=\Gamma/2$ and  partially entangled initial state $p=5/6$. 
	}
\label{fig:esd}
\end{figure}

In \autoref{fig:esd}, we show the peculiar behaviour of the four considered quantum correlations: CHSH-nonlocality, $\rm F_3$-steering, usefulness for teleportation and entanglement, for the partially entangled initial state $p=5/6$. These four quantum correlations exhibit the phenomenon of sudden death as shown in \autoref{fig:esd}(b), in particular, the entanglement-based correlations lie down to zero before $\sim0.4\Gamma t$. From all of these, only entanglement exhibits revival after a certain time, as shown in \autoref{fig:esd}(c). 	
In contrast, \autoref{fig:esd}(a) shows that the hidden versions of all the aforementioned  quantum correlations reach non-zero values and, most surprisingly, that such values achieve the maximal amount allowed by quantum theory, here represented by the value of $1$. 

In words, this result is telling us that although the considered quantum correlations naturally die after some finite time and never naturally rebirth again, they can still be extracted by means of local filtering operations and furthermore, that they can be taken to the maximal amount allowed by quantum theory. It is also interesting to note that besides local filtering operations (or SLOCCs), no other mechanism is being considered in this scenario. The system is naturally evolving to states which display the phenomenon of maximal hidden quantum correlations.

Whilst the physical system is indeed reaching maximal hidden correlations, one can argue that the effect of the local filters is actually helping the system to \emph{recover} the initial correlations which were lost by decoherence, rather than helping the system to extract correlations in a raw \emph{emergence} of hidden quantum correlations. In order to further explore this phenomenon, it is then natural to consider scenarios where the physical system does not possess quantum correlations at the beginning of the dynamics, as discussed in the next section.

\subsection{Sudden birth of maximal hidden quantum correlations}

\begin{figure}[h!]
	\centerline{
		\includegraphics[scale=0.8]{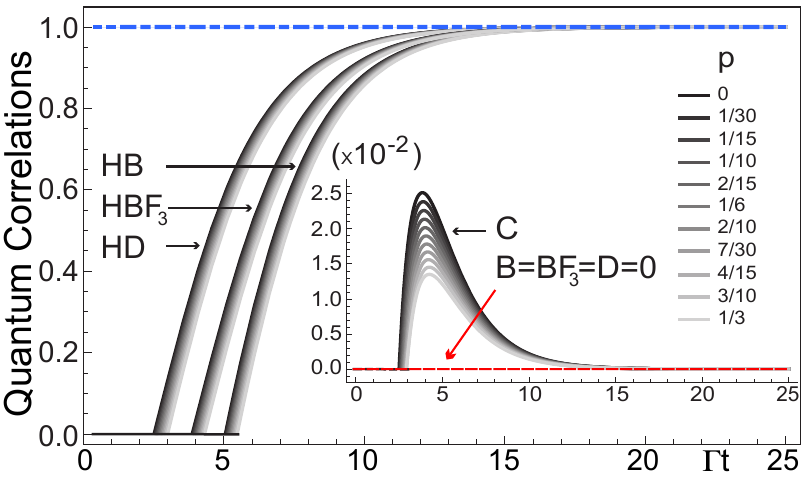}
	}
	\caption{
	Hidden quantum correlations: hidden CHSH-nonlocality (HB), hidden $\rm F_3$-steering ($\rm HBF_3$), and hidden usefulness for teleportation (HD). The dynamics portray  $\gamma=\Gamma/2$ and initial separable states $p\in[0,1/3]$. The inset shows the sudden birth of entanglement (concurrence) achieving values of order $10^{-2}$, and the unfiltered entanglement-based quantum correlations not being present (red dashed line at zero), meaning that, for all $t$: ${\rm B}(\rho(t))=0$ (CHSH-nonlocality), ${\rm BF_3}(\rho(t))=0$ ($\rm F_3$-steering), and ${\rm D}(\rho(t))=0$ (usefulness for teleportation). Notwithstanding this, the filtered correlations achieve a sudden birth of maximal hidden quantum correlations (see the horizontal blue dashed line at $1.0$, as a guide to the eye).
	}
\label{fig:fig2}
\end{figure}

Entanglement-based quantum correlations at a collective damping $\gamma=\Gamma/2$ are plotted in \autoref{fig:fig2}. The inset shows the sudden birth of concurrence. It has been done for ten steps of the initial (separable) state parameter $0\leq p\leq1/3$. The entanglement sudden birth phenomenon takes place (even though $p=0$; the maximally mixed state). Concurrence achieves a small maximum value (up to $\sim2.5\times10^{-2}$) and decays asymptotically to zero. Surprisingly, this weak entanglement behaviour however, still allows for the maximisation of entanglement-based properties of the system by means of local filtering; hidden usefulness for teleportation, hidden $\rm F_3$-steering and hidden CHSH-nonlocality suddenly appear and reach their maximum value ($\text{HB} \rightarrow \text{HBF}_3 \rightarrow{\rm HD}\rightarrow 1$) even though they  were all \emph{completely zero} along the whole dynamics before the local filtering. This result holds for all the spectrum of $p\in[0,1]$. In particular, for $p>1/3$, after the concurrence exhibits the sudden death and revival phenomena, all the entanglement-based hidden quantum correlations asymptotically achieve their maximum (as in \autoref{fig:esd} for $p=5/6$). As shown in \autoref{fig:fig2}, the sudden birth takes place at different times for distinct correlations, in complete agreement with the well known hierarchy amongst these correlations \cite{criterion_UfT}, with hidden CHSH-nonlocality exhibiting the more delayed sudden birth phenomenon.

This extreme hidden phenomenon holds for any $\gamma=k\Gamma$ with $k\neq0$ (similar results are obtained for negative values of $\gamma$). The particular case $\gamma=\Gamma$ corresponds to qubits very close to each other and is not a desirable physical scenario for exploring spatially-separated correlations. The case $k\rightarrow0$ on the other hand, is a very interesting scenario allowing enhanced correlations in long distance-separated qubits. An important fact in the range $k\in(0,1)$ where the maximisation takes place, is that the time of the sudden birth phenomenon for the hidden correlations strongly depends on the value of $\gamma$. Hence, the stronger the collective damping, the earlier the correlations appear in the dynamics (not shown). The specific case $k=0$, which can be thought of as two particles interacting with independent reservoirs, is not considered here because the only way to get maximal correlations by filtering occurs for initial maximally entangled states, as previously shown in \cite{isasi1}; the reason being that the Bell-diagonal state recovered after filtering coincides with the same initial entangled state. 

An interesting scenario to explore this maximal hidden correlation phenomenon is the case of a quantum system undergoing a non-Markovian dynamics. As it is well-known, death and revival behaviour of entanglement naturally arise  as a consequence of non-Markovianity~(see e.g., \cite{bellomo}). A discussion of hidden quantum correlations within a non-Markovian evolution (under both collective and independent qubit coupling to a bath) is part of an upcoming work \cite{DSR2022}. As some preliminary results in this regard, we have explored hidden correlations arising from the Jaynes-Cumming model in terms of the Markovian to non-Markovian transition parameter. As expected, local filtering enhances the correlations but we did not find a scenario in which the maximal phenomenon takes place. On the other hand, a question that arises from our results is whether different collective effects lead to similar phenomena like the one presented here. Following the Markovian dynamics here described, we consider two additional processes that exhibit collective effects: \emph{absorption} and \emph{pure dephasing}. For the sake of completeness, we have examined the behavior of hidden quantum correlations in both cases on the same set of initial states. Our findings are as follows: being absorption an energy-perturbing process, the maximal hidden correlation phenomenon takes place in the same way as the emission process here assumed. Again, the collective absorption parameter is the responsible for the phenomenon to occur. In contrast, the entanglement-based correlations are not optimised under the presence of the collective pure dephasing. This can be partially understood from the structure of the Lindblad super-operator of the master equation, which involves only the diagonal $\sigma_z$ Pauli operator. The solution of the master equation is such that two of the diagonal elements (here denoted by $a(t)$ and $d(t)$) remain unchanged by the dephasing dynamics. Hence, as explained before, the collective emission/absorption is crucial for the maximal phenomena to occur. This then evidences the importance of exploring distinct physical processes to investigate relevant resources as quantum correlations. Additionally, the introduction of a unified framework for quantifying both standard and hidden quantum correlations is also part of an upcoming work~\cite{DSR2022}.

To summarise, this result establishes that after a sudden birth of entanglement by a small amount, all the entanglement-based hidden quantum correlations do also exhibit sudden birth and, more interestingly, they all achieve their maximum values. It is worth stressing that all entanglement-based standard correlations are identically zero during the whole dynamics. Overall, we can therefore interpret this as the system displaying an ``all from nothing" behaviour; even though the physical system started with zero correlations, its natural evolution together with appropriate local filters allowed the system to be transformed to states close to displaying maximal quantum correlations. This is corroborated by means of the following unsteerability analysis of the evolved states during the sudden birth dynamics. 

\subsection{EPR-unsteerability and Bell-locality}

\begin{figure}[h!]
	\centering
	\includegraphics[scale=0.7]{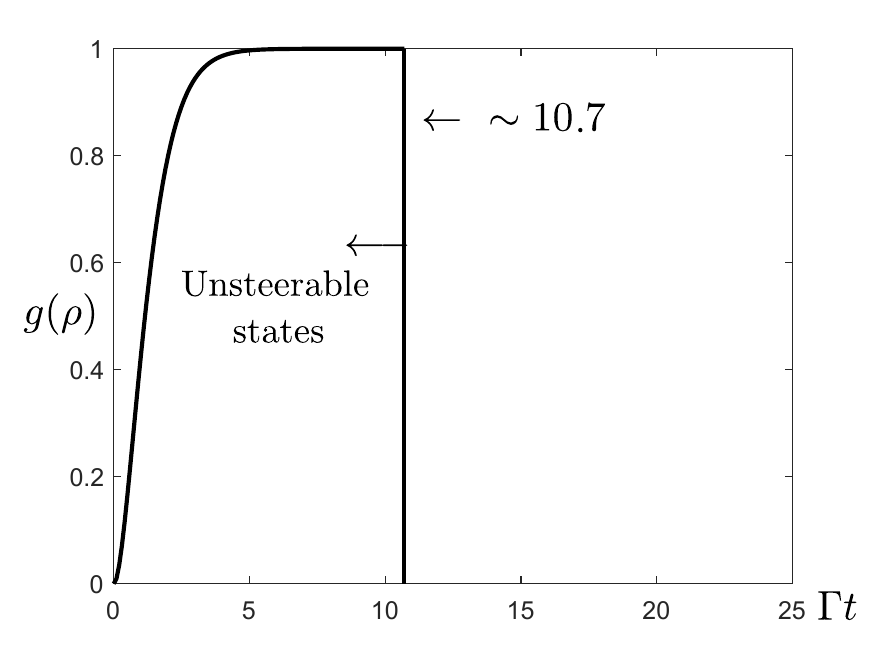}
	\vspace{-0.5cm}
	\caption{Unsteerability sufficient criterion for the dynamics in \autoref{fig:fig2} of the main text $(p=0)$. The function in \autoref{eq:g} against the temporal evolution.}
	\label{fig:fig3}
\end{figure}

Given an arbitrary two-qubit state $\rho \in D(\mathds{C}^2\otimes \mathds{C}^2)$ which can be represented by its $R$ matrix or, equivalently, by the three quantities $(\bm{a}, \bm{b}, T)$, as in \autoref{eq:R}, we consider its associated canonical form \cite{QSE2014}, which is a new state with the quantities $(\bm{c}, \bm{d}=\bm{0}, \tilde T)$ given by \cite{QSE2014}: 
\begin{eqnarray}
	\bm{c}=&
	\gamma_b^2(\bm{a}-T\bm{b}),\hspace{0.4cm}  
	\gamma_b \coloneqq \frac{1}{\sqrt{1-b^2}}, \hspace{0.2cm}
	b=|\bm{b}|, \label{eq:QSEcentre}\\
	\tilde T=&
	\gamma_b 
	\left(
	T-\bm{a}\bm{b}^T
	\right)
	\left(
	\mathds{1}+
	\frac{\gamma_b-1}{b^2}
	\bm{b} \,\bm{b}^T
	\right).
	\label{eq:QSEmatrix}
\end{eqnarray} 
With this canonical form, we have the following EPR-unsteerability criterion \cite{UN3}. If the state $\rho$ satisfies the inequality:
\begin{equation}
	g(\rho)
	\coloneqq 
	\max_{\bm{x}} 
	\left[
	\left(
		\bm{c} \cdot \bm{x} 
	\right)^2
	+
	2 ||\tilde T \bm{x}||
	\right]
	\leq 1,
	\label{eq:g}
\end{equation}
with $\bm{x} \in \mathds{R}^3$, $||\bm{x}||=1$, the standard Euclidean norm, then, the state is EPR-unsteerable for arbitrary projective measurements \cite{UN3}. In \autoref{fig:fig3} we address this sufficient criterion for the dynamics considered in \autoref{fig:fig2} $(p=0)$.

In \autoref{fig:fig3} we see that this sufficient criterion detects entangled states admitting an LHS model (the states display sudden birth of entanglement after around $\Gamma t \sim 2.5$). The criterion however stops working around $\Gamma t \sim 10.7$ (meaning $g(\rho)>1$), after which it cannot be used to address unsteerability. Therefore, the states in the region $2.5 \lesssim \Gamma t \lesssim 10.7$ are displaying hidden EPR-steering. Taking into account that the existence of LHS models imply the existence of LHV models, this result together with the CHSH-nonlocality behaviour shown in \autoref{fig:fig2}, demonstrate that the set of states up to $\Gamma t \sim 10.7$ transfom from local states to Bell-nonlocal states. A more elaborate numerical approach \cite{UN1, UN2} could be implemented in order to try to extend the region for which the existence of LHS models is guaranteed, this, beyond that of the criterion in \autoref{eq:g}. Here, it would also be interesting to explore whether it is possible to achieve a similar result in terms of \emph{genuine} hidden quantum correlations, in the sense of the unfiltered dynamics admitting LHS and LHV models for general POVMs (beyond projective measurements).

With the spirit of operationally justify the appeal of having a physical system naturally tailored to display maximal hidden quantum correlations, a discussion on the filtering probability is addressed in the next section.

\subsection{Trade-off between the sudden birth of maximal hidden quantum correlations and its filtering probability}

\begin{figure}[h!]
    \centering
    \includegraphics[scale=0.7]{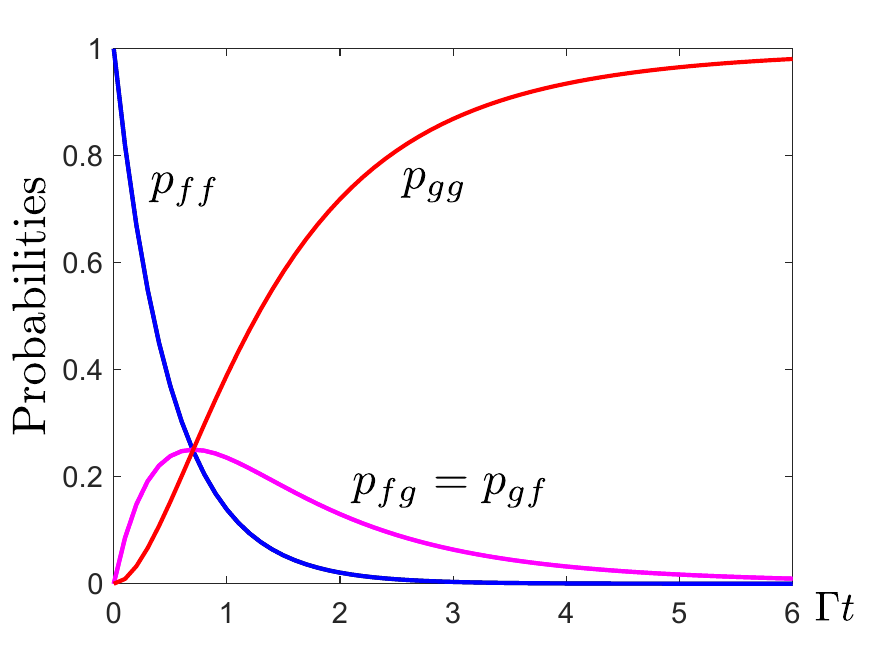}
    \vspace{-0.5cm}
    \caption{
    Filtering probabilities against the temporal evolution of the system. There is an interesting trade-off between the amount of quantum correlations which can be extracted (\autoref{fig:fig2}) and the filtering probability with which this can be done ($p_{ff}$). 
    }
    \label{fig:fig4}
\end{figure}

In order to debate the probability with which the filtering process occurs, we need to derive the KLM-SLOCC for the states of interest. In general, calculating the KLM-SLOCC is not a trivial task, luckily however, the system under consideration here is always evolving through states in the X-form and therefore, this simplifies the derivation of the KLM-SLOCC as follows. The KLM-SLOCC for two-qubit X-form states is given by \cite{xslocc2005, xslocc2020}:
\begin{eqnarray}
    f_A
    \coloneqq
    \begin{pmatrix} 
	\left( dc/ab \right)^\frac{1}{4} & 0\\
	0 & 1
	\end{pmatrix}
	,
    \hspace{0.5cm}
	f_B
    \coloneqq
    \begin{pmatrix} 
	\left( db/ac \right)^\frac{1}{4} & 0 \\
	0 & 1
	\end{pmatrix} .
\end{eqnarray}
We can then calculate the probability of the filtered state $\rho_{ff}$ as:
\begin{equation}
    p_{ff}={\rm Tr}\left[ \left( f_A^ \dagger f_A \otimes f_B^\dagger f_B \right) \rho_X \right]
    =
    2d \left[ 1 + \left( \frac{bc}{ad} \right)^\frac{1}{2} \right] ,
\end{equation}
and similarly for the probabilities $p_{fg}, p_{gf}, p_{gg}$. With these tools at hand, we can now explore the filtering probabilities for the dynamics corresponding to the sudden birth phenomenon, and this is plotted in \autoref{fig:fig4}.

In \autoref{fig:fig4} we see that whilst the amount of quantum correlations achieve their maximum value $\rightarrow 1$ as time increases (as per \autoref{fig:fig2}), the filtering probability with which such procedure is achieved tends to the minimum value $p_{ff}\rightarrow 0$, although being strictly greater than zero. In words, this is telling us that Nature does not give away maximal hidden quantum correlations for free. The extraction of these maximal correlations is possible, with a filtering probability strictly greater than zero $p_{ff}>0$, but they quickly become difficult to achieve $p_{ff}(t\rightarrow \infty) \rightarrow 0$.

For a better understanding of our finding, some algebra on the density matrix can be made as follows; still in the specific sudden birth scenario shown in \autoref{fig:fig2}, the asymptotic behavior of the density matrix can be written as $\rho(\epsilon(t))=\epsilon(t) \ketbra{\psi^-}{\psi^-} + (1-\epsilon(t)) \ketbra{00}{00}$, with the singlet state $\ket{\psi^-} = \frac{1}{\sqrt{2}} (\ket{01}-\ket{10})$, and the coefficient satisfying $\epsilon(t\rightarrow +\infty) \rightarrow 0$. In this case, the optimal local filters for Alice and Bob can be defined as $f_A=f_B=\epsilon \ketbra{0}{0}+\ketbra{1}{1}$, and consequently the filtered state becomes $\rho'(\epsilon)=\frac{1}{1+\epsilon-\epsilon^2}[(\epsilon-\epsilon^2) \ketbra{00}{00}+\psi^-]$, which effectively satisfies $\rho'(\epsilon(t\rightarrow +\infty) \rightarrow 0)\rightarrow \psi^-$, and in turn it naturally maximises the quantum correlations of interest. However, analysing the filtering probability, on the other hand, we get $\epsilon^3(1+\epsilon-\epsilon^2)$, which rapidly decays to zero when $t\rightarrow +\infty$ and therefore, evidences the interesting trade-off between the amount of correlations that can be extracted, and the rate at which it could happen.

From a theoretical point of view, this trade-off therefore imposes a fundamental limit to the the amount of correlations which can be extracted by means of local filtering operations. In addition to this theoretical interest, the results here presented determine, in a practical manner, the amount of resources that should be invested (in terms of the filtering probability) in order to extract such maximal amount of hidden quantum correlations.

\section{Conclusions}
\label{sec:conclusion}

We have shown that physical systems undergoing a Markovian dissipative quantum dynamics can exhibit the phenomena of death-revival and sudden birth of \emph{maximal hidden quantum correlations}. We have considered atoms or quantum emitters undergoing a sudden birth of entanglement (of order $10^{-2}$, measured in terms of the concurrence), which \emph{do not} allow the birth of entanglement-based correlations in the form of: CHSH-inequality violation for Bell-nonlocality, $\rm F_3$-inequality violation for EPR-steering, and usefulness for teleportation. Furthermore, we have reported a region for this sudden birth of entanglement dynamics which even admits a description in terms of LHS and LHV models for projective measurements and consequently, ruling out the presence of either EPR-steering or Bell-nonlocality (this, beyond not violating the particular CHSH and $\rm F_3$ inequalities here considered). Notwithstanding all of these undesirable characteristics however, we have shown that this apparently not-so-useful dynamics is actually in possession of quantum correlations,  hidden from us, but that can be revealed by means of local filtering operations. Specifically, that at any point of the dynamics, there exist local filters that transform the state into a new state which now exhibits all these three forms of correlations. Most surprisingly however, that these hidden quantum correlations can be enhanced to the maximal amount allowed by quantum theory, i.e., those of the maximally entangled pure two-qubit state (the singlet, up to local unitaries). 

Regarding the physical system here considered, the presence of the collective damping parameter is crucial for the maximisation of quantum correlations to occur. It is also worth noting that scenarios with long-distance separated quantum emitters are allowed as well, due to it is not mandatory to invoke the dipolar interaction. For example,  in the case of plasmonic waveguides, inter-emitter distances equal to or greater than the operational wavelength ($\sim 1$ $\mu$m)~\cite{SR2014} could be considered. Furthermore, we notice that another physical scenario as it is the case of collective absorption also exhibits the maximal hidden correlation phenomenon here reported.

Despite the immediate appeal of the physical system displaying such hidden quantum correlations, we have also shown however, that there also exists a \emph{trade-off} between the amount of these hidden quantum correlations which can be extracted, and the filtering probability with which such protocol can be implemented, therefore establishing a fundamental limit for the extraction of quantum correlations by means of local filtering operations. Explicitly, the more hidden quantum correlations Alice and Bob want to extract out of the system, the lower the filtering probability of the protocol. From a practical point of view on the other hand, these results determine the amount of resources (in terms of the filtering probability) which should be invested in order to extract such maximal hidden quantum correlations.

\section*{Acknowledgements}

A.F.D. thanks P. Skrzypczyk for helpful discussions, and acknowledges financial support from COLCIENCIAS (Grant 756-2016) and the UK EPSRC (EP/L015730/1). C.E.S. acknowledges funding from Universidad de C\'ordoba (Grants CA-097; FCB-14-17 and FCB-08-19). J.H.R. is grateful to Universidad del Valle for support (Grant CI~1269). We also acknowledge funding from the Colombian Science, Technology and Innovation Fund--General Royalties System (Fondo CTeI---Sistema General de Regal\'ias) and Gobernaci\'on del Valle del Cauca (Grant BPIN 2013000100007).

\bibliographystyle{IEEEtran}
\bibliography{bibliography.bib}
\end{document}